\documentclass[a4paper,fleqn,usenatbib]{mnras}

\usepackage{mathptmx}
\usepackage{txfonts}

\usepackage[T1]{fontenc}
\usepackage{ae,aecompl}

\usepackage{graphicx}	

\newcommand{\se}[1]{\S\ref{sec:#1}}

\newcommand{\Fig}[1]{Figure~\ref{fig:#1}}

\newcommand{\be}{\begin{equation}}
\newcommand{\ee}{\end{equation}}
\newcommand{\bea}{\begin{eqnarray}}
\newcommand{\eea}{\end{eqnarray}}

\newcommand{\msun}{{\rm M}_\odot}
\newcommand{\Msun}{M_\odot}

\newcommand{\ifm}[1]{\relax\ifmmode#1\else$\mathsurround=0pt #1$\fi}
\newcommand{\kms}{\ifmmode\,{\rm km}\,{\rm s}^{-1}\else km$\,$s$^{-1}$\fi}

\newcommand{\kpc}{\,{\rm kpc}}

\newcommand{\ltsima}{$\; \buildrel < \over \sim \;$}
\newcommand{\lsim}{\lower.5ex\hbox{\ltsima}}
\newcommand{\gtsima}{$\; \buildrel > \over \sim \;$}
\newcommand{\gsim}{\lower.5ex\hbox{\gtsima}}

\def\la{\langle}

\def\sy{\,M_\odot\, {\rm yr}^{-1}}

\def\cmc{\,{\rm cm}^{-3}}
\def\cms{\,{\rm cm}^{-2}}

\def\Mv{M_{\rm v}}

\def\Ms{M_*}

\usepackage{color}

\title[Disc Formation]{Formation and Settling of a Disc Galaxy During the Last 8 Billion Years in a Cosmological Simulation}

\author[Ceverino et al.]{
Daniel Ceverino,$^{1}$\thanks{E-mail: ceverino@uni-heidelberg.de}
Joel Primack$^2$,
Avishai Dekel$^3$,
Susan A. Kassin$^4$
\\
$^{1}$Zentrum für Astronomie der Universität Heidelberg, Institut für Theoretische Astrophysik, Albert-Ueberle-Str. 2, 69120 Heidelberg, Germany\\
$^{2}$Department of Physics, University of California, Santa Cruz, CA,
95064, USA \\
$^{3}$Center for Astrophysics and Planetary Science, Racah Institute of Physics, The Hebrew University, Jerusalem 91904, Israel \\
$^{4}$Space Telescope Science Institute, 3700 San Martin Drive, Baltimore, MD 21218, USA
}

\date{Accepted XXX. Received YYY; in original form ZZZ}

\pubyear{2016}

\begin{document}
\label{firstpage}
\pagerange{\pageref{firstpage}--\pageref{lastpage}}
\maketitle

\begin{abstract}
We present results of a high-resolution zoom cosmological simulation of the evolution of a low-mass galaxy with a maximum velocity of $V\simeq100 \ {\rm km}\,{\rm s}^{-1}$ at $z\simeq0$, using the initial conditions from the AGORA project \citep{Kim}.
The final disc-dominated galaxy is consistent with local disc scaling relations, such as the stellar-to-halo mass relation and the baryonic Tully-Fisher.
The galaxy evolves from a compact, dispersion-dominated galaxy into a rotation-dominated but dynamically hot disc in about 0.5 Gyr (from $z=1.4$ to $z=1.2$).
The disc dynamically cools down for the following 7 Gyr, as the gas velocity dispersion decreases over time, in agreement with observations.
The primary cause of this slow evolution of velocity dispersion in this low-mass galaxy is stellar feedback. It is related to the decline in gas fraction, and to the associated gravitational disk instability, as the 
disc slowly settles from a global Toomre $Q>1$ turbulent disc to a marginally unstable disc ($Q\simeq1$). 
\end{abstract}



\section{Introduction}

Any theory about the formation of galactic discs should reproduce not only the final ($z\simeq0$) global properties of disc galaxies and the observed scaling relations \citep{Verheijen01, Swaters00, Salim07, McGaugh05, Courteau07, Benson07, Reyes11, Simons15} 
but it should also explain the evolution of the disc kinematics, from the formation of a rotationally supported galaxy to its settling down as a dynamically cold disc.

Current studies roughly agree in a basic picture of the global kinematics of local ($z\simeq0$) disc galaxies.
A disc morphology is usually correlated with gas kinematics dominated by rotation.
This rotation follows the Tully Fisher relation between the rotational velocity ($V$) and the galaxy absolute magnitude, stellar or baryonic mass \citep[][and references therein]{TullyFisher77, McGaugh00, Courteau07}.
Therefore, the importance of disordered and/or turbulent motions, measured by the gas velocity dispersion ($\sigma$),  is currently low in disc galaxies, so that the present day
ratio of velocities is low, $\sigma / V=0.1-0.15$ \citep{Leroy09, Epinat10, Green14}.
Todays disc galaxies are rotationally supported and dynamically cold.
However, was this also true in the past?

Works on the evolution of the Tully-Fisher relation at high redshifts have shown an increase in the relative contribution of disordered motions \citep{Flores06, Kassin07, Vergani12}.
Measurements of $\sigma/V$ using different techniques and tracers give much higher values at $z>0$.
This means that rotationally dominated discs are dynamically hotter at higher redshifts \citep{Epinat09, Kassin12, Tacconi13,Wisnioski15}.
This leads to the concept of 'disc settling' \citep{Kassin12}.
In this scenario, discs cool down over time and settle in a dynamically cold disc.
However, there is little understanding of this process of settling. 
It is not clear how fast is that process and what are the main drivers in the evolution of gas kinematics with time.

The trend of higher $\sigma/V$ at higher redshifts continues to $z=2-3$, such that the majority of the high-z rotating galaxies are clumpy and dynamically hot discs \citep{Genzel06, Forster09, Gnerucci11}. 
Slightly lower mass galaxies at these high redshifts are often dispersion-dominated, where rotation is only a minor component in the gas kinematics \citep{Law09}.
Are these galaxies the progenitors of dynamically hot discs? In spite of recent progress, there are many unknowns about the formation of clumpy discs. One of the caveats is that
 current complete samples of high-z galaxies only contain massive, baryonic-dominated galaxies ($\Ms\ge 10^{10} \ \Msun$).
These are not the progenitors of normal, low-mass discs at $z\simeq0$.

The formation and evolution of low-mass discs with a final stellar mass of a few times $10^9 \ \Msun$ and maximum rotational velocities of $\sim$100 $\kms$ remains poorly understood.
This mass (or velocity) scale 
is relevant because it
marks the transition below which supernovae feedback is most efficient in the regulation of star formation at galactic scales \citep{DekelSilk86}.
As feedback remains one of the most poorly understood ingredients in galaxy formation models,
the evolution of disc kinematics at this mass scale can place important constrains to current models of feedback. 
For example, too strong feedback may yield 'puffy' and more turbulent discs than observed at a given redshifft \citep{Stinson06}.

Current cosmological simulations of galaxy formation at this mass scale, which corresponds to a halo mass of about $2 \times 10^{11} \ \Msun$ \citep{Moster13, Behroozi13}, have successfully matched many observations of local, low-mass disc galaxies \citep{Brooks11, Brook12, Hopkins14, Christensen14, Obreja14, Christensen16, Santos16}.
However, most of these previous works have not addressed the evolution of the disc kinematics from high redshifts to now.
Only \cite{Kassin14} discussed the evolution of ordered and disordered motions using the simulations first shown in \cite{Brooks11}.
The simulated values follow a trend similar to observations but their velocities are too high,
because their stellar masses are higher than in the observed sample.
However, their $\sigma/V \le0.1$ for cold gas at $z\simeq1$ is too low in disagreement with
MUSE observations of galaxies with a similar mass and redshift \citep{Contini16}.
There is also little discussion about the role of the different physical processes in the evolution of disc settling as well as in the formation of discs.

In this paper, we study the formation of a low-mass disc galaxy and the evolution of disc kinematics using a cosmological zoom-in simulation.
Section \se{run} provides details about the simulation.
We first compare the properties of the simulated galaxy at $z\simeq0$ with other simulations of similar mass and with observations of local galaxies (\se{z0comp}).
In section \se{formation} we follow the transformation from a compact, dispersion-dominated galaxy to a dynamically hot but rotation-dominated disc.
In section \se{settling} we address the settling of the galaxy into a dynamically cold disc by $z\simeq0$.
Finally section \se{conclusions} is devoted to the conclusions and discussion.

\section{The simulation}
\label{sec:run}

The initial conditions of this simulation corresponds to the proof-of-concept test of the AGORA project \citep{Kim}. The zoom-in, N-body-only cosmological simulation has a galactic halo of $\Mv =1.7 \times 10^{11} \Msun$ at $z=0$.
\footnote{$\Mv$ is defined as the mass enclosed in a sphere of mean density contrast $\Delta(z)$ \citep{BryanNorman98}.}
It has a  "quiescent" assembly history, having its last significant merger (1:5 mass ratio) at $z\sim1.5$.
This is an ideal laboratory to study the formation and settling of a galactic disc during the following $\sim$7 Gyr after the merger in a fully cosmological setup.
The simulation has a minimum DM mass of 
$3.5 \times 10^4 \ \msun$, while the particles representing single stellar populations that were formed in the simulation
have a minimum mass of $10^3 \ \msun$. 
The maximum spatial resolution is 163 comoving pc.

%
The N-body+Hydro simulation was performed with the  \textsc{ART} code
\citep{Kravtsov97,Kravtsov03}, which accurately follows the evolution of a
gravitating N-body system and the Eulerian gas dynamics using an AMR approach.
Beyond gravity and hydrodynamics, the code incorporates 
many of the physical processes relevant for galaxy formation.  
These processes, representing subgrid 
physics, include gas cooling by atomic hydrogen and helium, metal and molecular 
hydrogen cooling, photoionisation heating by a constant cosmological UV background with partial 
self-shielding, star formation and feedback, as described in 
\citet{Ceverino09}, \citet{CDB}, and \citet{Ceverino14}. 
Star formation is assumed to occur at densities above a threshold of 1 $\cmc$ and at temperatures below $10^4$ K.  The code implements a stochastic star formation model that yields the empirical Kennicutt-Schmidt law \citep{Kennicutt98}, as shown in \cite{Ceverino14}.

In addition to thermal-energy feedback, the simulation uses radiative feedback.
This model, called RadPre\_LS\_IR in  \citet{Ceverino14}, adds a non-thermal pressure, radiation pressure, to the total gas pressure in regions where ionising photons from massive stars are produced and trapped. 
In short, the radiation pressure is proportional to the ionising luminosity of a single star particle.
This pressure is added if the cell (and their closest neighbours) contains a star particle younger than 5 Myr and whose column density exceeds $10^{21}  \cms$.
Photoionization and photoheating are also included in these regions by using tabulated results from \textsc{CLOUDY} \citep{Ferland98}.
Finally, the model also includes a  moderate trapping of infrared photons, only if the gas density in the host cell exceeds a threshold of 300 $\cmc$.

This model differs from other recent implementations of feedback. 
it goes beyond the thermal-only feedback \citep{Stinson06, Stinson13, Keller15,  Schaye15} and it does not shutdown cooling in the star-forming regions.
It does not impose a wind solution \citep{Vogelsberger14}, so that outflows are generated in a self-consistent way \citep{Ceverino16}.
Our implementation is more similar to the radiative feedback model in \cite{Agertz15}.
Within our model, both radiative and supernovae feedback act in concert and they are equally important. The combination of early feedback from radiation and late feedback from supernovae regulates star formation within galaxies \citep{Ceverino14}.

\section{Comparison with other simulations and observations at z=0}
\label{sec:z0comp}

Before assessing the formation and settling of the galactic disc in this simulation, we are going to compare the properties of the simulated galaxy at the last snapshot at a redshift of $z=0.1$ with available observations and other simulations of disc-dominated galaxies with a similar halo mass.

\begin{figure}
	\includegraphics[width=\columnwidth]{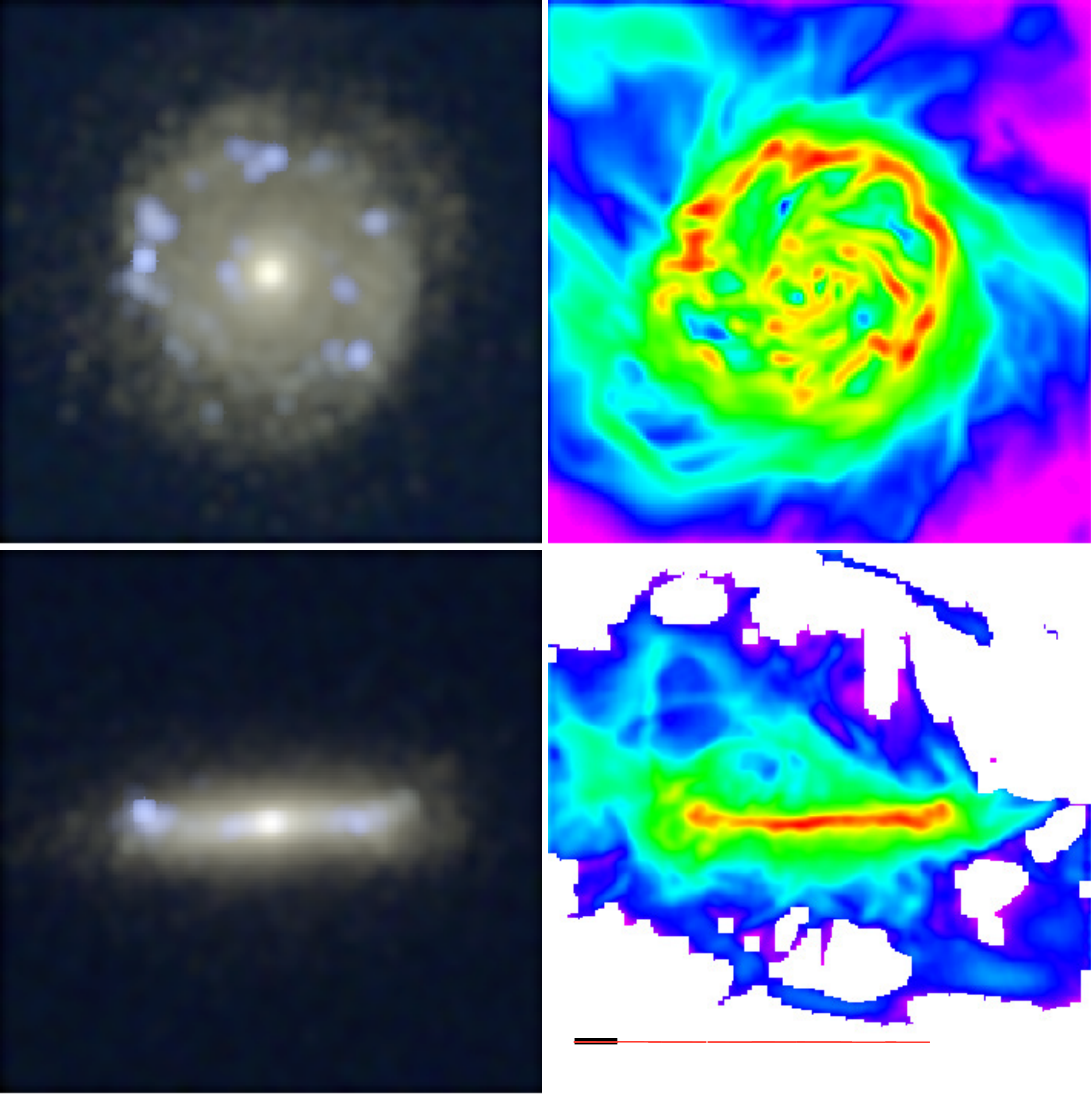}
    \caption{UBV stellar light (left) and gas (right) of the disc galaxy at $z=0.1$ in the face-on (top) and edge-on (bottom) views (40x40 kpc), according to the direction of the angular momentum of the cold ($T<10^4 K$) gas. The horizontal bar in the bottom-right panel represents a length of 3 kpc. }
    \label{fig:z0}
\end{figure}

The galaxy shows a disc morphology (\Fig{z0}) both in stellar light and gas.
The disc extends to 10 kpc with a total stellar mass of $\Ms=3.2 \times 10^9 \Msun$. This is the regime of low-mass late-type galaxies.
The gas looks clumpy and flocculent, similar to the HI distribution of galaxies with similar mass \citep{Kim98}.
The star-forming regions appear in the U-band and they coincide with large gas complexes, mostly organised in a star-forming ring.
The star-formation-rate, $SFR=0.2 \sy$ is consistent with observations of star-forming galaxies with similar mass in the local Universe \citep{Salim07, Avila11}. 

\begin{figure}
	\includegraphics[width=\columnwidth]{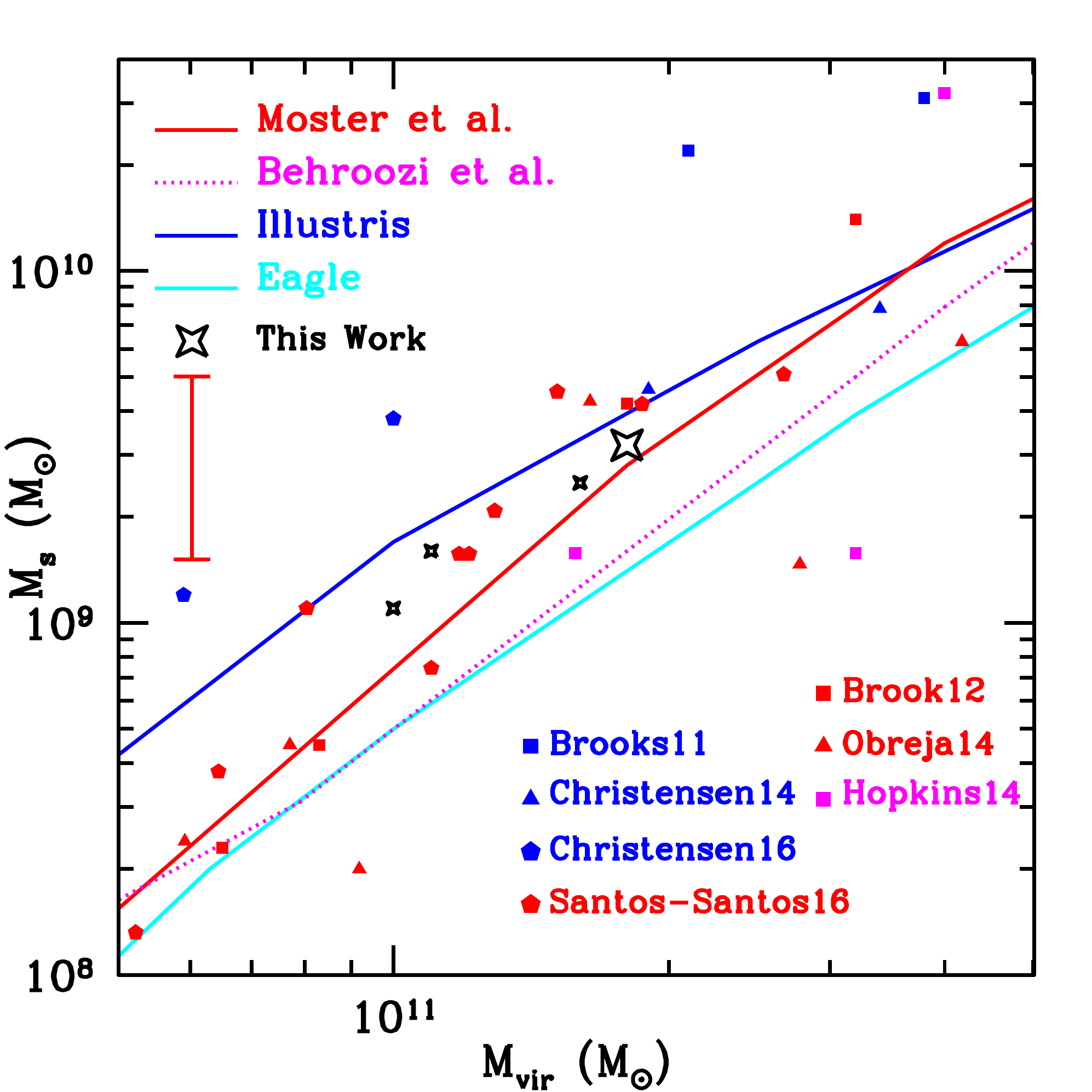}
    \caption{Stellar mass versus virial mass of the simulation (black stars) in comparison with results from abundance matching and other simulations at z=0.  
    Our results at z=1.5, 1, 0.5 and z=0.1 (from left to right) agree reasonable well
     with previous simulations and abundance matching models, within the typical uncertainty at these mass scales, shown as a vertical error bar. }
    \label{fig:fS}
\end{figure}

The stellar-to-halo mass ratio, or galaxy efficiency, is consistent with current abundance matching models \citep{Moster13,Behroozi13}, within the typical uncertainties (\Fig{fS}).
It is also consistent with the galaxy efficiency in the Illustris simulation \citep{Vogelsberger14} and slightly higher than the median ratio in the Eagle simulation \citep{Schaye15}, although within the scatter.
Other recent zoom simulations agree relatively well with our results, despite the different codes and feedback models implemented \citep{Brook12, Hopkins14, Obreja14, Christensen14, Christensen16, Santos16}.
Older simulations, like \cite{Brooks11}, show a greater discrepancy, with a stellar-to-halo mass ratio 10 times higher than our study.

\begin{figure}
	\includegraphics[width=\columnwidth]{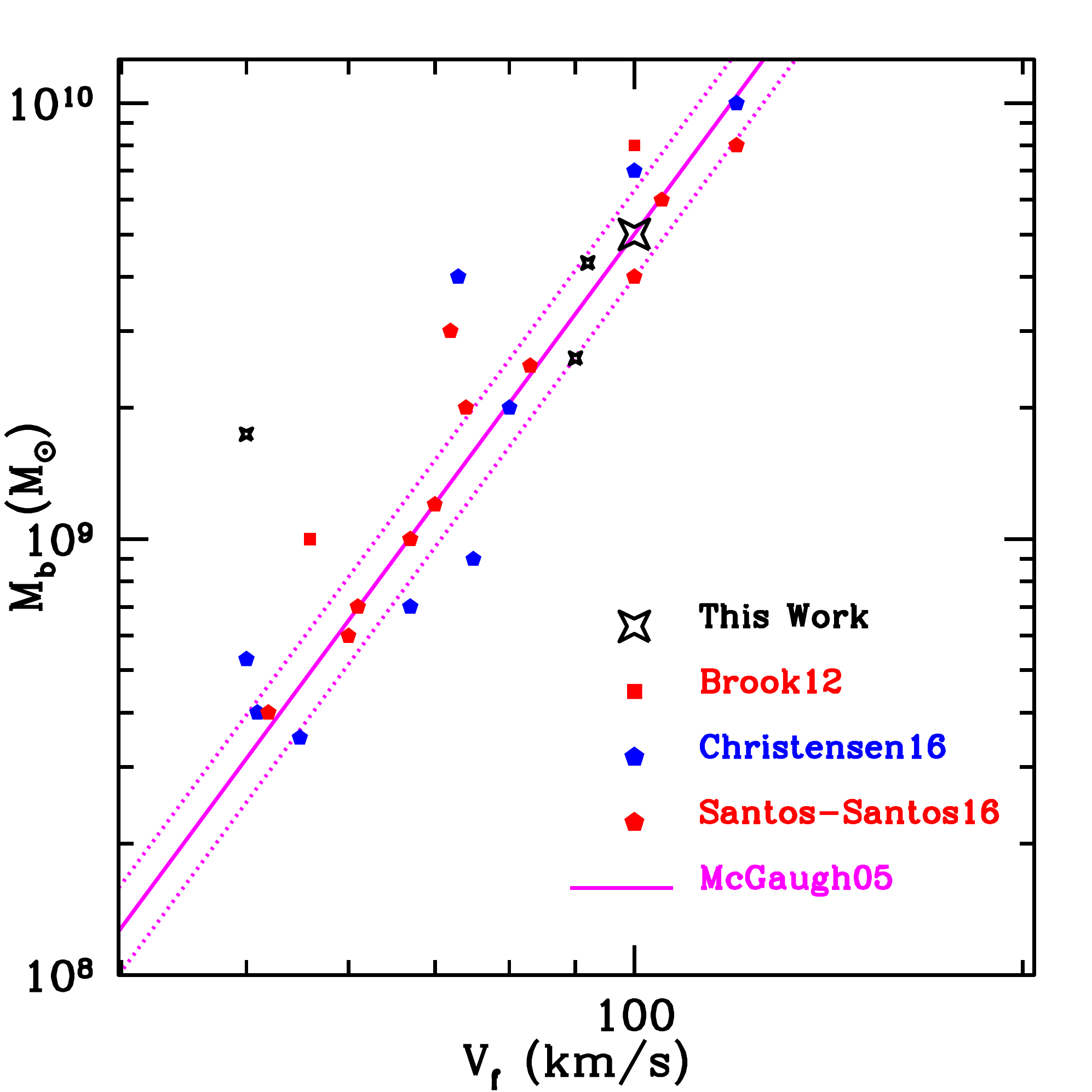}
    \caption{Baryonic Tully-Fisher relation and comparison with observations and other simulations at z=0. The simulation 
    (black stars as in \Fig{fS}) agrees remarkably well with the observed relation 
    at $z\simeq0$ \citep{McGaugh05}.}
    \label{fig:TF}
\end{figure}

The simulation at z=0.1 also agrees remarkably well with the baryonic Tully-Fisher relation \citep{McGaugh05} (\Fig{TF}).
The flat rotation velocity is estimated as the rotational velocity of the cold gas ($T<10^4 K$) at 2.2 disc scalelengths (8.4 kpc).
This good agreement is not completely unexpected because the other recent zoom simulations also match this scaling relation,
although they measure it  in different ways. \cite{Brook12} use 3.5 scalelengths, while \cite{Christensen16} and \cite{Santos16} use the asymptotic velocity 
However, it is important to note that these results are based on circular velocity profiles, rather than on the actual rotational velocity of the gas 
As we will see in the next sections, the rotational velocity could be significantly lower than the expected circular velocity if turbulent or non-rotational motions, driven by gravity or feedback, are significant.
This effect could lower some of these results, specially if strong feedback maintains large irregular motions.

\begin{figure}
	\includegraphics[width=\columnwidth]{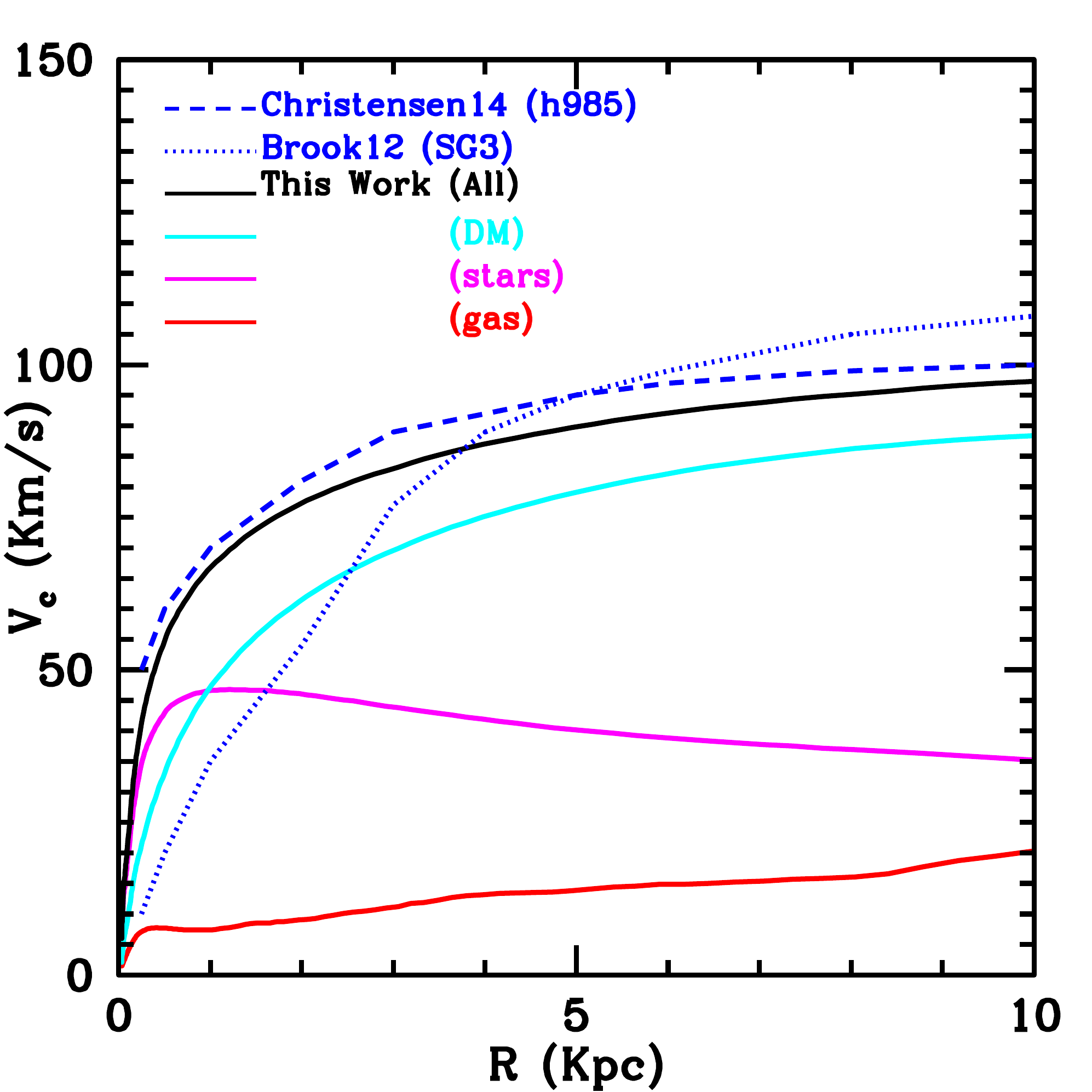}
    \caption{Circular velocity profile and comparison with other simulations of similar halo mass. The profile shows a rise with radius, mostly dominated by DM.}
    \label{fig:Vc}
\end{figure}

The circular velocity profile, $V_c= ( GM(r<R)/R )^{0.5}$, shows a rise with radius, mostly dominated by DM outside the inner first kpc (\Fig{Vc}). 
The profile reaches an asymptotic velocity of about 100 km/s, very similar to the rotation velocity at large radii.
The simulation avoids the excessive concentration of stellar mass, reflected in the lack of a central peak in the velocity profile that has plagued simulations of galaxy formation for many years \citep{Abadi03, Mayer08}.
More recent simulations also show this rise in galaxies with a similar halo mass ($\Mv \simeq10^{11} \msun$). For example,  \Fig{Vc}  compares our results with the h985 galaxy in \cite{Christensen14} and with the SG3 run in \cite{Brook12}.
There is a good agreement at large radii. This means that they have roughly the same mass within $R\simeq 10 {\rm kpc}$.
 However, we found some disagreements at smaller radii. The total profile from SG3 is significantly steeper at $R< 4 {\rm kpc}$. This is probably the result of the strong and explosive feedback model used in \cite{Brook12} that not only reduces the central baryonic concentration, but also expands the dark matter halo, generating a significant reduction of the total mass in its inner part. Due to the lack of published rotational velocity profiles that could be compared with observations, it is difficult to address the validity of this approach.

\begin{figure}
	\includegraphics[width=\columnwidth]{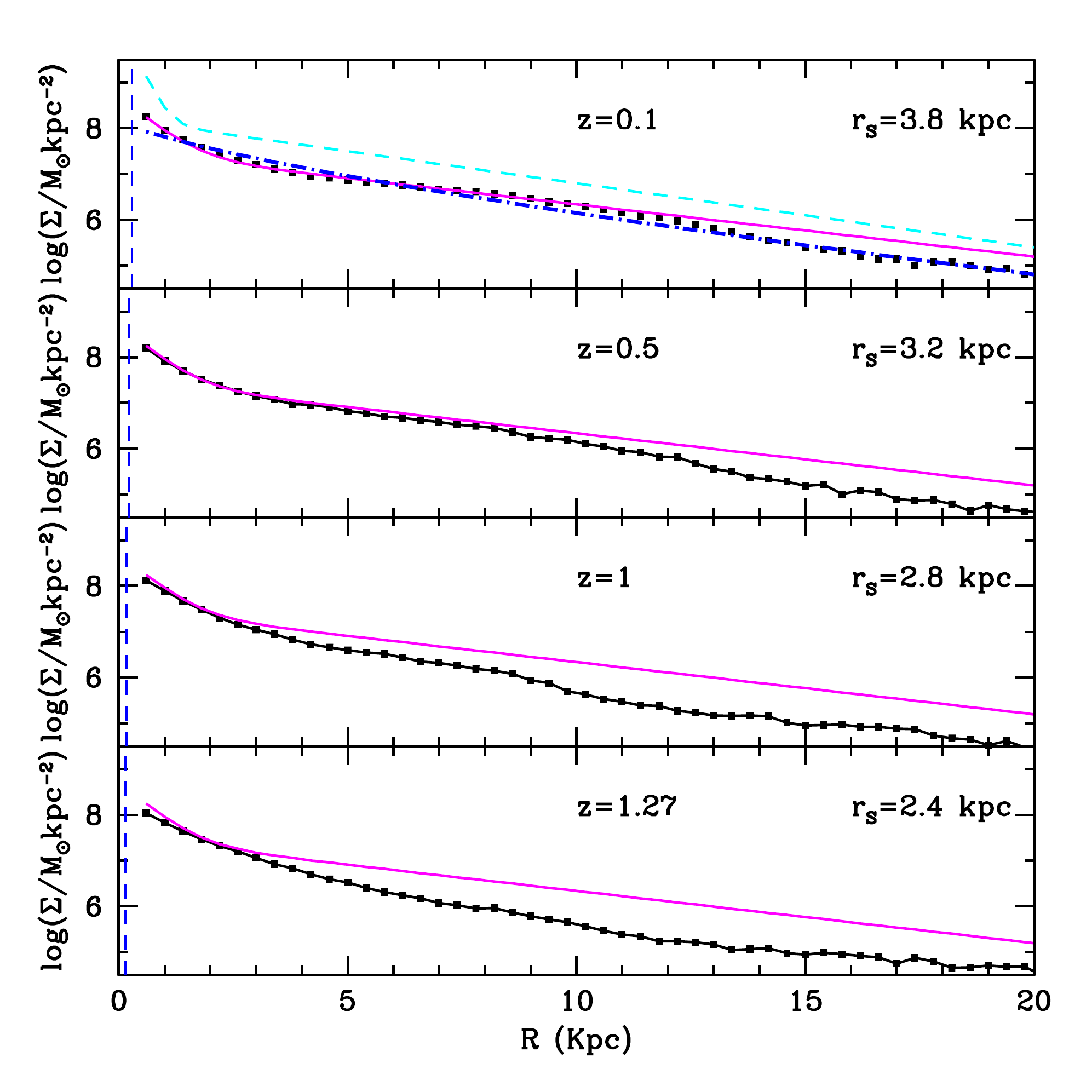}
    \caption{Stellar mass surface density at different redshifts. The dash-dotted blue line shows a single-component Sersic fit with a index of n=1.4 and the solid magenta lines in all panels show a double-exponential fit at z=0.1 $r_s$ is the disc scalelength at each redshift. Only 20\% of the stellar mass is concentrated in the inner component. Therefore the simulated galaxy is dominated by an extended disc at $z\simeq0$.
    The dashed cyan line shows the too massive h985 run from Brooks et al. (2011). The vertical dashed blue line shows the force resolution.  }
    \label{fig:SigmaS}
\end{figure}

The top panel of \Fig{SigmaS} shows the profile of stellar mass surface density at $z\simeq0$.
A single component Sersic fit gives an index of $n=1.4$, close to an exponential profile, and an effective radius of 4.1 kpc. 
However, there is a clear up-turn in the profile inside the first kpc, so two exponential components provide a better fit. The resulted disc-only scale-lenght is 3.8 kpc, consistent with the observed size-velocity relation \citep{Courteau07}.

As a comparison, \Fig{SigmaS} also shows the profile of the h985 run discussed in \cite{Brooks11}.
That profile shows a much higher surface density, consistent with being a massive disc with an excessively high stellar-to-halo mass ratio.
Our run shows a less massive disc plus a relatively smaller central component with a mass of only 20\% of the total stellar mass, consistent with estimations from local SDSS galaxies \citep{Benson07}. 
The fraction of light in the B-band is even smaller. Only 14\% of the luminosity in the B-band is coming from the first kpc.
Therefore, we conclude that the simulated galaxy is a low-mass, disc-dominated galaxy at $\simeq0$.

\section{Disc Formation}
\label{sec:formation}

In this section, we address the transformation from a compact, dispersion-dominated galaxy after the last significant merger at $z=1.5$ to a dynamically hot but rotation-dominated disc galaxy at lower redshifts ($z\simeq1$).

\subsection{From a compact galaxy to an extended thick disc}

\begin{figure}
	\includegraphics[width=0.99 \columnwidth]{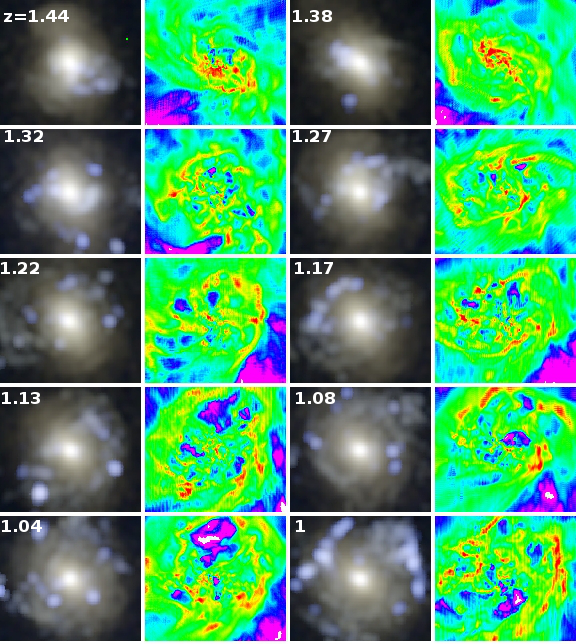}
		\includegraphics[width=0.99 \columnwidth]{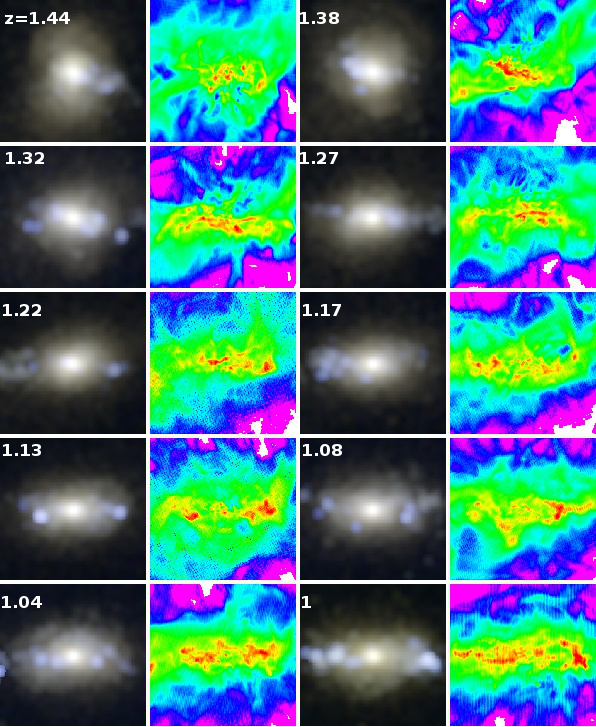}

    \caption{UBV stellar light (left) and gas (right) at different redshifts. 
    Each panel is 20x20 kpc$^2$. 
    The top five rows show the face-on views and the rest show the edge-on views. 
    The first two snapshots ($z=1.44-1.38$) show a compact galaxy with most of the gas and star formation at the centre.
    The second row ($z=1.32-1.27$) shows  a more distributed gas. An extended but thick stellar disc  grows slowly with time.}
    \label{fig:disc}
\end{figure}

\Fig{disc} shows the temporal evolution of the stellar light and gas distribution at different redshifts after the minor merger at $z=1.5$.
In the first two snapshots ($z=1.44-1.38$), the galaxy is compact, with an effective radius of about 1.5-2 kpc. 
A large fraction of gas is accumulated at the centre, where most of the star-forming clouds (in red in the gas distribution) are concentrated.
This concentration of gas is due to the inflow of material in a compaction event triggered by the minor merger \citep{Zolotov15}.
The event results in a compact galaxy both in rest-frame U and V bands.

After compaction, fresh gas comes along smooth gas streams and builds an extended gaseous disc. 
In the second pair of snapshots ($z=1.32-1.27$), star-forming clouds are preferentially found at large radii. There is little gas at the centre because the inflow of gas within the disc has declined \citep{Zolotov15} and now star-formation happens preferentially in the disc \citep{DekelBurkert}.
The rest-frame U light is significantly more extended than the rest-frame V band, which tracks the bulk of the stellar distribution, mostly concentrated at the centre.
This is shown in the density profile at the bottom panel of \Fig{SigmaS}.
Therefore, the signature of this new-born disc is a negative gradient in the U-V color, if spatially resolved \citep{Liu16}. 

After this short period of first star-formation in a gaseous disc, an extended disc can also be seen in the V band at $z\le1.22$. The star-formation continues throughout the gaseous disc up to a radius of 10 kpc at z=1. These gaseous and stellar discs are thick, as shown in the edge-on views. This reflects a dynamically hot kinematics.

\subsection{From a dispersion-dominated galaxy to a rotation-dominated disc}

\begin{figure}
	\includegraphics[width=\columnwidth]{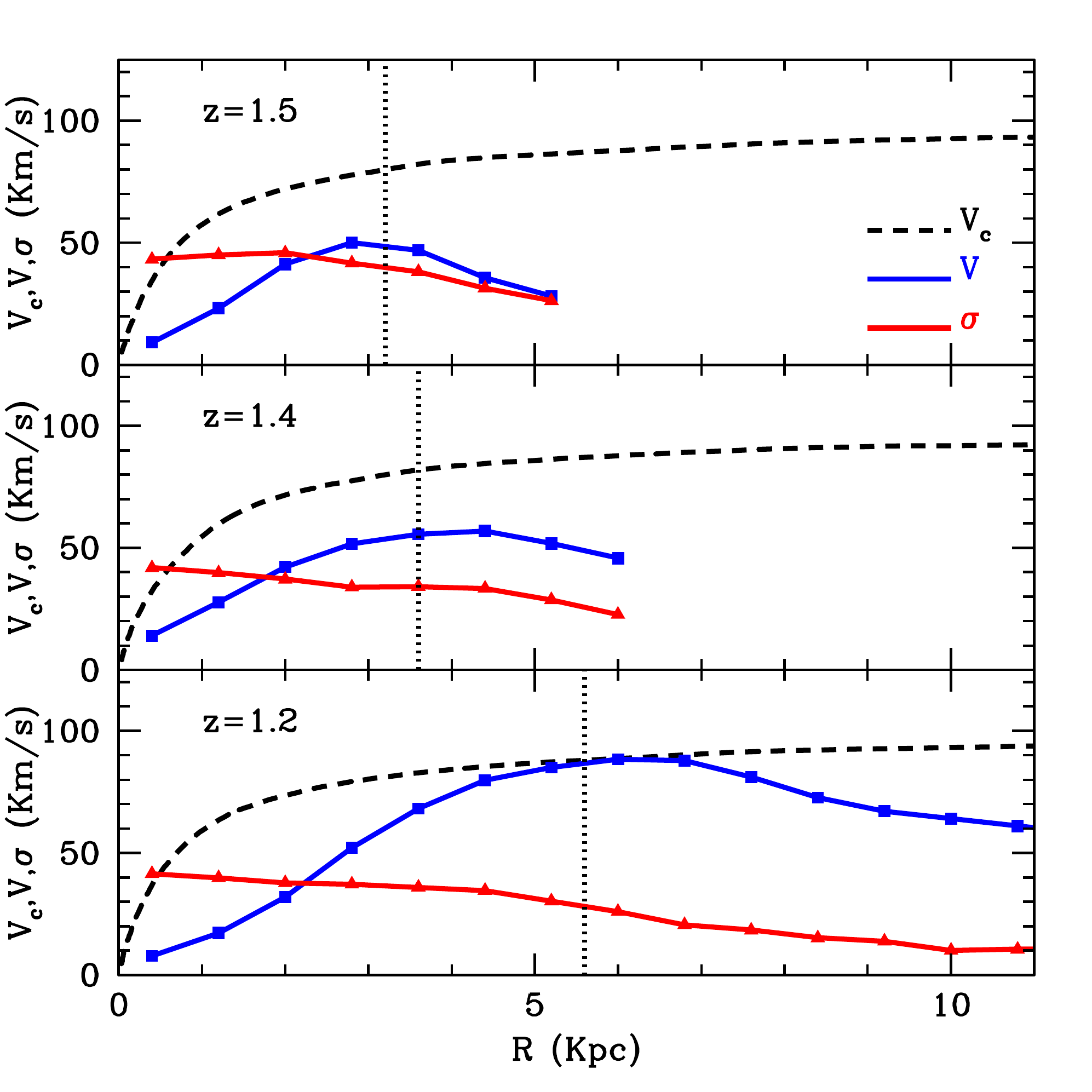}
    \caption{Evolution of the profiles of circular velocity ($V_c$), rotational velocity ($V$) of gas, and its dispersion ($\sigma$) during the formation of the disc. The vertical dotted lines mark 2$r_e$.
    The galaxy evolves from a dispersion-dominated and compact galaxy to a rotation-dominated but dynamically hot disc.}
    \label{fig:kinematics}
\end{figure}

During the compaction phase ($z\sim1.5$), the gas kinematics is dominated by velocity dispersion, rather than rotation (\Fig{kinematics}).  
A good measure of the degree of rotation support is the {\it rotation parameter}  \citep{Ceverino12},
 defined as the contribution of rotation to the support against gravitational collapse in the Jeans equation, ${\mathcal R}=V^2/V_c^2$, where the rotational velocity ($V$) and the circular velocity ($V_c$) are measured at twice the effective radius.
In the compaction phase, rotation only accounts for 40\% of the support at $2r_e$.
Therefore, a high velocity dispersion of about 40 \kms dominates over rotation.
This value is about constant at all radii inside $2r_e$.
This is an example of dispersion-dominated ($\sigma/V  \geq 1$) galaxies that 
account for the majority of the galaxies at low masses ($\Ms \leq 10^{10} \Msun$) and high redshifts \citep{Law09, Newman13}.

A rotationally supported disc starts to grow after the compaction event ($z\le1.4$).
The velocity dispersion is still high but it decreases with radius from 40 \kms at the inner radius to 30 \kms at $2r_e$.
The rotational velocity increases with radius and it accounts for 50\% of the support at that radius (or $\sigma/V\sim 0.6$) at $z=1.4$.

The rotational support grows with time upto ${\mathcal R}\sim 1$ at $z=1.2$. This means that rotation contributes 100\% against gravitational collapse at large radii.
Velocity dispersion at that radii is still high ($\sigma=30$ kms) but it is significantly lower than the rotational velocity ($\sigma/V\sim 0.4$). 
The disc is rotationally supported but kinematically hot.
Dispersion still dominates the kinematics at low radii, such that the rotational velocity is much lower than the circular velocity inside the effective radius.
Most of the gas is distributed in a turbulent and clumpy disc, which resembles a low-mass analog of the massive discs observed at redshifts $z=1-2$  \citep{Genzel06, Forster09, Wisnioski15}.

Another interesting feature of the rotation curves of these early discs is the strong decrease of rotation velocity at even larger radii ($r>2r_e$). 
At $z=1.2$, the rotation curve has declined by 20\% at 3$r_e$ with respect to its maximum value at about 2$r_e$. 
This decline may be due to significant non-rotational motions at the edge of the gaseous discs, where smooth gaseous streams deliver fresh gas into the disc \citep{Danovich15}.
Alternatively, gas at these large radii may not be in dynamical equilibrium and still flowing into the galaxy. 
Future studies will clarify the nature of the truncation of these early discs.

\section{Disc Settling}
\label{sec:settling}

\begin{figure}
	\includegraphics[width=\columnwidth]{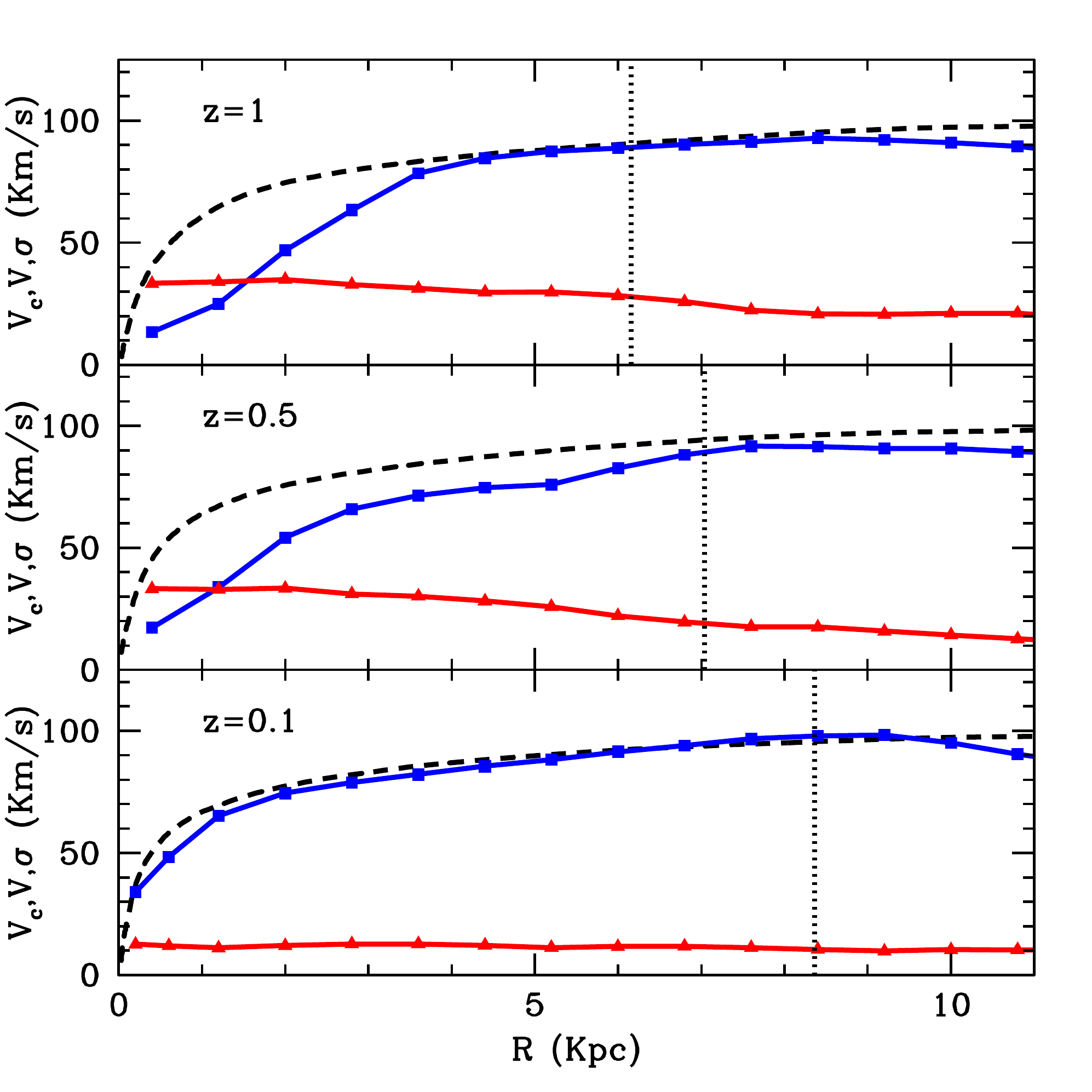}
    \caption{Evolution of the profiles of circular velocity ($V_c$), rotational velocity ($V$) of gas, and its dispersion ($\sigma$)  during the settling of the disc. The labels are the same as in \Fig{kinematics} with the exception of the vertical dotted lines, which now mark 2.2$r_s$. }
    \label{fig:kinematics2}
\end{figure}

From the formation of a rotationally-supported disc at $z\sim1.3$ to the last snapshot at $z=0.1$, the stellar mass inside 15 kpc grows by a factor $\sim$2, mostly within the disc, according to the evolution of the profiles of stellar surface density (\Fig{SigmaS}).
The mass within the  innermost kpc remains mostly unchanged since the last compaction event.
The outer regions significantly grow in mass as the disc grows inside-out. 
The disc scale-length from a double-exponential fit grows from $r_s=2.4 \kpc$ to $r_s=3.8 \kpc$ between $z=1.27$ and $z=0.1$.

Due to the fresh supply of gas, the gas mass remains approximately constant within 15 kpc ($M_{\rm Gas}=2 \times 10^9 \Msun$)
Therefore, the gas fraction decreases from 60\% of all the baryons at $z=1.3$ to 40\% at $z=0.1$.
Finally, the total mass, mostly dominated by dark matter, remains approximately constant within that radius ($M(r<15 \kpc)=3 \times 10^{10} \Msun$).
Therefore, the total circular velocity profile is mostly unchanged since $z\simeq1$.

\Fig{kinematics2} shows the evolution of the gas kinematics since $z=1$.
The disc remains rotationally  supported ($V\simeq V_c$) at large radii ($r\simeq 2.2 r_s$) for almost 7 Gyr (from $z=1.2$ to $z=0.1$).
The radius at which rotation gives full support (${\mathcal R}\sim 1$) decreases with time, from $r>4 \kpc$ at $z=1$ to $r>0.5 \kpc$ at $z=0.1$.
It seems that the disc tends to dynamically cool from the outer to the inner regions.
This is due to the decrease of the spatially-averaged velocity dispersion with time.
It gradually decreases  from $\sigma=30 \kms$ at $z=1$ to $\sigma\simeq10 \kms$ at $z=0.1$.

\begin{figure}
	\includegraphics[width=\columnwidth]{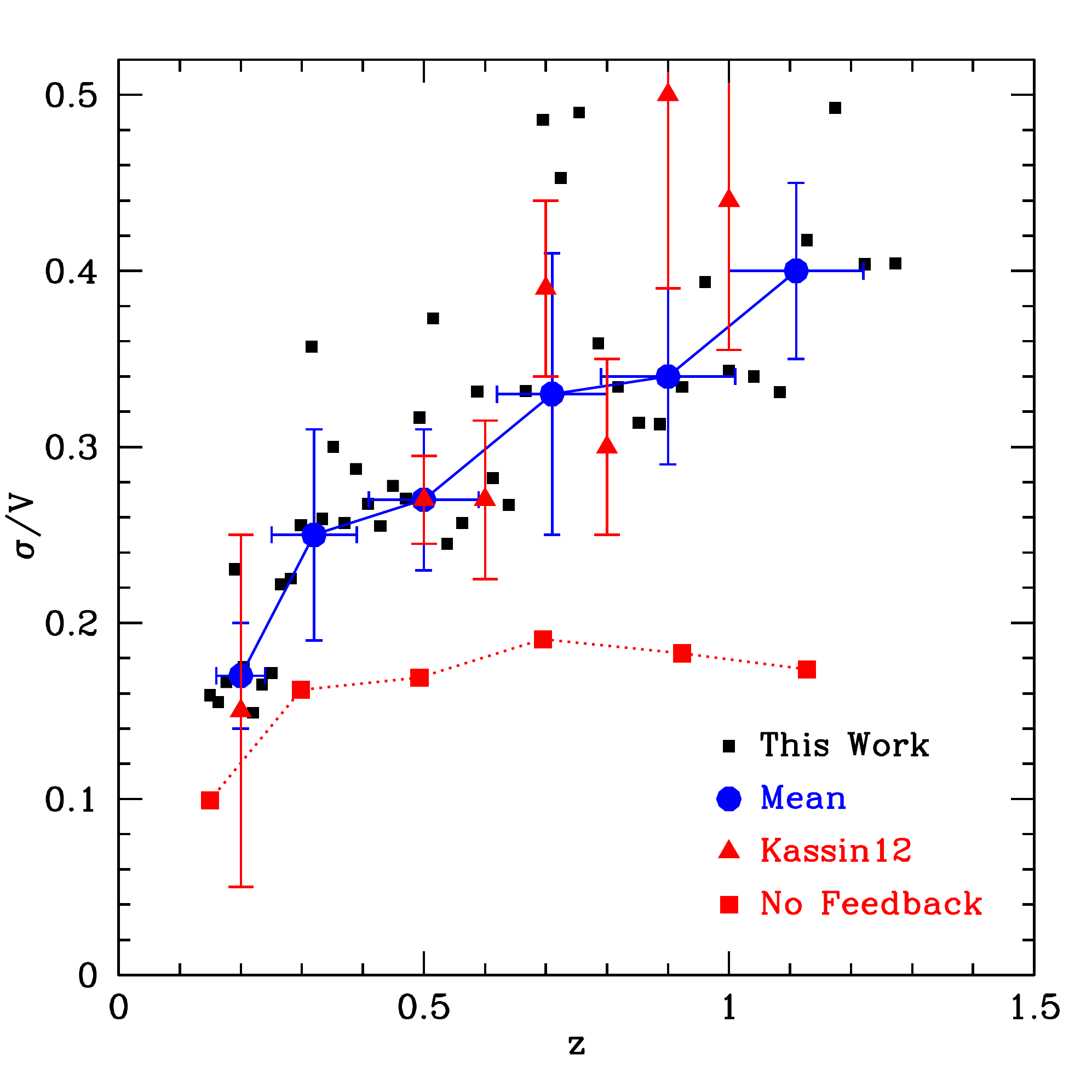}
    \caption{Gradual decline of $\sigma/V$ with decreasing redshift (solid line), consistent with observations. The simulation without feedback (dotted line) shows very little evolution.}
    \label{fig:sigmaV}
\end{figure}

During 7 Gyr of evolution, the disc dynamically cools and settles in a state of low velocity dispersion with close-to circular orbits.
\Fig{sigmaV} shows the gradual settling of the disc due to the decrease of the $\sigma /V$ ratio. 
This is the ratio between the median velocity dispersion in the radial bins inside 2.2$r_s$ and the rotational velocity at $2.2r_s$, described above. 
We check the velocity dispersion for gas at different temperatures ($T<10^3$ K and $T<10^4$ K) and they both give the same results.
This slow decline is consistent with the observations of disc settling in discs of similar mass and redshifts \citep{Kassin12}.
Some tension may exist at the highest redshifts observed ($z\simeq1$), in which the observed values of $\sigma /V\simeq0.4-0.5$ are slightly higher than the values coming from simulations $\sigma /V \simeq0.3-0.4$.

\subsection{What process controls the decrease in velocity dispersion?}
\label{sec:dispersion}

Several processes may control or regulate the velocity dispersion in galactic discs.
Discs are self-gravitating and rotating objects subject to gravitational instabilities \citep[][and references therein]{DSC}. 
Gravitational energy is transferred to turbulence and other non-circular motions by gravitational torques \citep{KrumholzBurkert10, Forbes14}.
Feedback from star-formation processes is another common source of turbulence and other bulk disordered motions in discs \citep{Thompson05, OstrikerShetty11, FaucherGiguere13}.
External processes like gas accretion or mergers may also contribute to keep some level of turbulence \citep{Genel12},
although this may not be important in the evolution of this particular galaxy because of its quiet accretion history.

It is difficult to disentangle the contributions from different internal processes, like instabilities or feedback, because both effects are present in self-gravitating discs.
In order to remove the short-term driver due to feedback, we perform the following experiments. At different times, we rerun the simulation without feedback for 
 $\Delta (1+z)^{-1}=0.03$ ( $\Delta$t=0.5 Gyr at $z=1$ and 0.4 Gyr at $z=0.1$).
Turbulence in discs decays in few dynamical times, so most of the velocity dispersion due to feedback processes within the disc have been dissipated after that period of time, when we compute again the velocity dispersion as done above. 
 In this way, we keep the long term effects of self-regulation of star-formation and remove the short-term source of turbulence due to feedback.
 We found a significantly lower dispersion ($\sigma\simeq15 \kms$) with little evolution over most of the redshift range.
This is the expected level of dispersion if turbulence is driven only by disc instabilities in a low-mass, marginally-unstable disc \citep{DSC,CDB, Cacciato12}. 
 At this mass scale, the DM halo mostly dominates the gravitational potential within the galaxy and this tends to stabilise the disc and maintains a low velocity dispersion. 
 The key variable controlling the degree of dispersion for a marginally-unstable disc  is the disc-to-total mass ratio within the galaxy radius ($\delta\simeq0.1$). This low value gives $\sigma /V\simeq0.1$ \citep[see Eq. 5 in][]{CDB}.
This low ratio is in clear contradiction with observations of high-z galaxies.
 Therefore, we conclude that most of the velocity dispersion observed in low-mass discs ($V\simeq100 \kms$) at high redshifts is driven by feedback from star formation processes.
 
 \begin{figure}
	\includegraphics[width=\columnwidth]{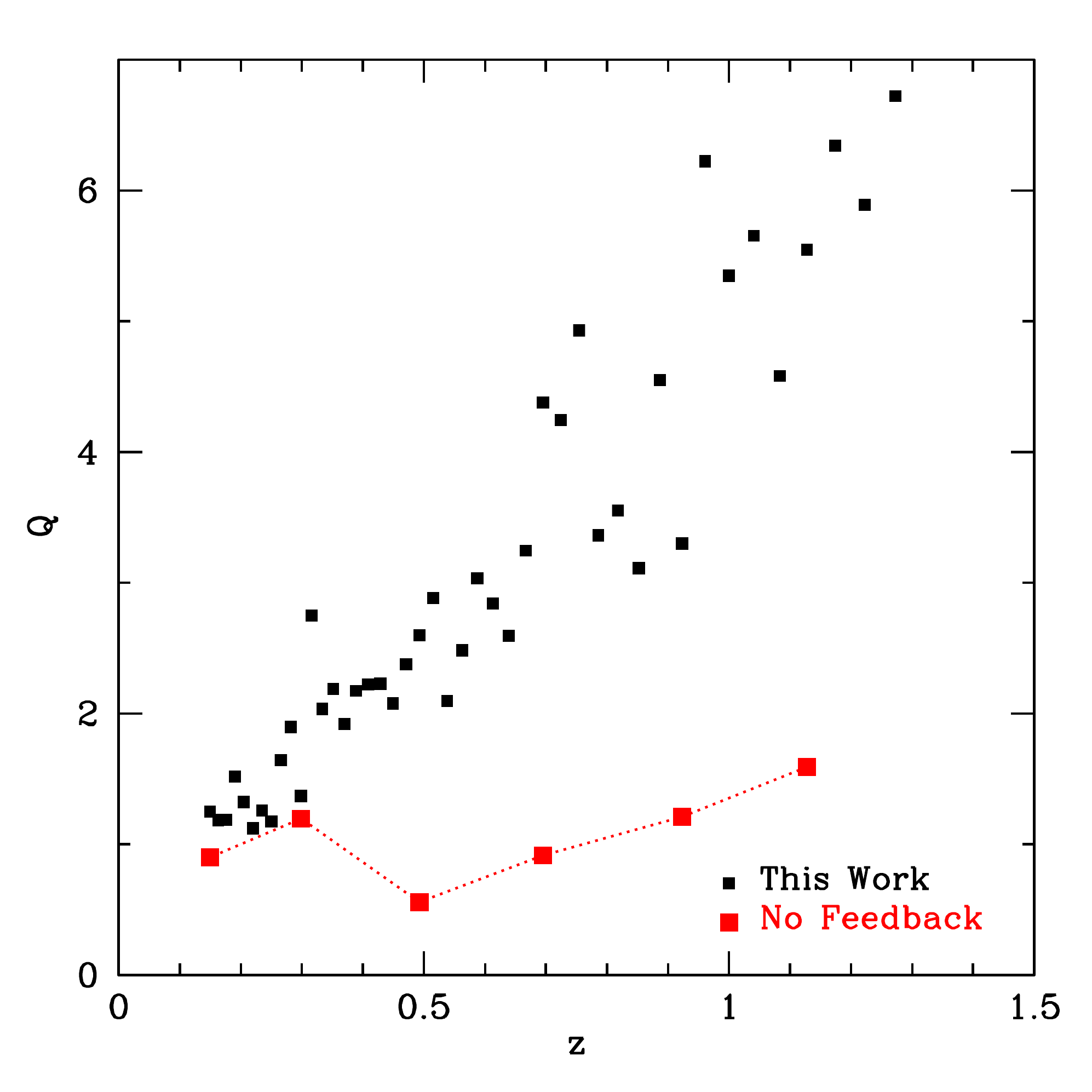}
	    \caption{Evolution of the Toomre parameter, Q, from a stable disc at high-z to a marginally unstable ($Q\simeq1$) disc at low redshifts. The run without feedback maintains the disc in the marginally unstable case, where disordered motions are driven only by gravity. }
	      \label{fig:Q}
\end{figure}  

\begin{figure}
	\includegraphics[width=\columnwidth]{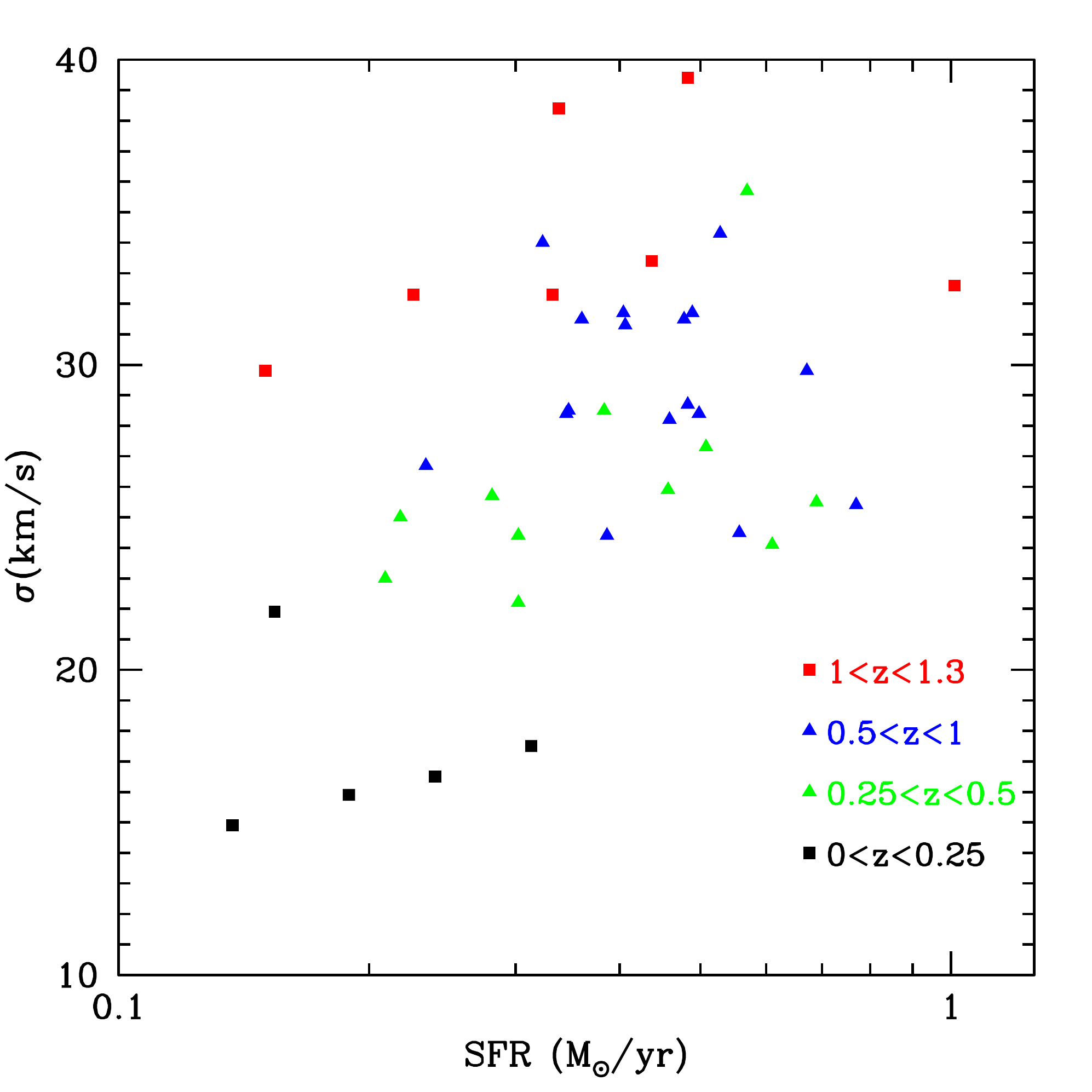}
    \caption{Velocity dispersion versus SFR at different redshifts. The velocity dispersion decreases due to the decrease of the gas surface density (and gas fraction) at lower redshifts.}
    \label{fig:sigmaSFR}
\end{figure}
 
 With the exception of the gas velocity dispersion, the global properties of the disc remain roughly similar after $\sim$500 Myr without feedback. 
 Only the SFR is significantly higher (a factor of 5 higher on average), due to the shutdown of the feedback-driven regulation. Therefore, the stellar mass is $\sim$50\% higher when compared with the feedback run after 500 Myr.
 However, the gas mass within the galaxy decreases only by $\sim$10\% because new accreting gas replenishes the gas consumed in stars.
 As a result, the gas fraction decreases only by $\sim$20\% averaged over all snapshots.
 Without feedback, the disc becomes slightly more massive, dynamically cold, and marginally-unstable.
 
The feedback-driven velocity dispersion stabilises the disc above the \cite{Toomre} $Q\sim1$ on kpc scales. 
 We measure the total Toomre Q parameter for both gas and stars using the \citep{WangSilk94} approximation described in \cite{Krumholz16},
 \begin{equation} 
 Q \approx \sqrt{2} \frac{ V \sigma f_{\rm g} } { \pi G  \Sigma R}
 \end{equation}
  where $f_{\rm g}$ is the gas fraction and $\Sigma$ is the disc surface density. 
  We add a velocity dispersion of 10 km/s in quadrature to the value coming from simulations in order to account for thermal motions within diffuse gas. 
  At each snapshot, we measure $Q$ at the same radius, $R=5 \kpc$, where $V\simeq 80 \kms$ at $z<1.3$ (\Fig{Q}). The estimations of $Q$ at different radius give a similar behaviour.  
We clearly see a slow decrease of $Q$ from high values, $Q\simeq6$ at $z\simeq1$ to $Q\simeq1$ at $z=0.1$.
In the runs without feedback, the disc always remains around the self-regulated, marginally unstable case ($Q\simeq1$), because gravity is the only driver of disordered motions.
With feedback, there is a slow evolution towards this self-regulated case, as the stellar disc grows and the gas surface density (and $f_{\rm g}$) decreases by a factor of 2 between $z=1$ and $z=0.1$. If feedback drives high velocity dispersions, we expect lower values associated with lower SFRs at lower redshifts (lower gas fractions).
 \Fig{sigmaSFR} shows this correlation but the scatter is large at a given redshift bin, consistent with observations \citep{Forster09}.

This feedback-driven scenario differs from standard stationary models of feedback-regulated star formation \citep{FaucherGiguere13}, where $Q$ is regulated at a fixed value of a few. In our scenario of DM-dominated discs, star-formation proceeds even if the global, kpc-averaged $Q$ is high,
most probably due to gravitational instabilities at smaller scales \citep{Inoue16}.
It seems that disc instability, star formation and feedback work in concert and they generate high velocity dispersion
but it is not clear how these processes interact with each other in the regime of weak self-gravity in DM-dominated discs.
Future work will focus on this complex interplay.
 



\section{Conclusions and Discussion}
\label{sec:conclusions}

We have performed a high-resolution cosmological zoom simulation of the formation and evolution of a low-mass galaxy with maximum velocity of 100 $\kms$ until $z\simeq0$, using the initial conditions from the AGORA project \citep{Kim}.
The main conclusions can be summarised as follows
\begin{itemize}
\item
The properties of the disc-dominated galaxy agree well with local disc scaling relations, such as the baryonic Tully-Fischer, as well as the stellar-to-halo mass ratio coming from abundance matching models and other simulations of similar halo mass.
\item 
The galaxy undergoes a compaction event triggered by a minor merger at $z=1.5$ \citep{DekelBurkert, Zolotov15}.
This event generates a compact, dispersion-dominated galaxy with a large fraction of its gas and star formation concentrated at the galaxy centre. 
\item
After this last compaction event, a rotation-dominated disc starts to grow, so that the rotational velocity is equal to the circular velocity at a radius of 2.2$r_e$ after 0.5 Gyr of evolution ($z\simeq1.2$).
\item
The new-born disc remains thick and dynamically hot, such that the velocity dispersion still dominates the dynamical support at small radii.
\item
The disc dynamically cools down for the following 7 Gyr, as the velocity dispersion (and $\sigma/V$) decreases over time, in agreement with observations \citep{Kassin12}.
\item
Simulations without feedback do not show this decrease. 
A scenario of feedback-driven turbulence explains this decrease by a similar decrease in the gas surface density (and gas fraction).
\item
The disc is globally Toomre stable ($Q\simeq6$) at $z\simeq1$ and it slowly evolves towards the marginally unstable case ($Q\simeq1$), as the stellar disc grows.
\end{itemize}

In section \se{dispersion}, we discuss the low level of velocity dispersion driven only by gravitational instabilities in DM-dominated galaxies.
This result cannot be extended to more massive discs, where baryons dominates the mass within the galaxy.
In that case, the disc-to-total mass ratio is much higher and the disc self-gravity drives violent disc instability \citep[VDI,][]{DSC, CDB}.
VDI maintains a self-regulated turbulent and clumpy disc with a much higher velocity dispersion, 
 $\sigma\simeq50 \kms$ \citep{Ceverino12}, as observed using integral field spectroscopy (IFS) \citep{Wisnioski15}.
In future work, we will perform mock IFS observations of 2D gas kinematics.

These results of disc settling cannot be extended to galaxies of similar mass at higher redshifts ($z>2$).
In that regime, the DM halo also dominates the mass within the galaxy, but its shape is highly prolate \citep{Allgood06}.
As a result of strong torques, an axial-symmetric disc is not able to form.
Instead, galaxies form with an elongated, triaxial or prolate shape \citep{Ceverino15b, Tomassetti16}.

The properties of the simulated galaxies are still sensitive to the assumptions about feedback processes.
The simulations reported here assume a moderate radiative feedback, with only a modest trapping of infrared photons.
Other simulations of halos with a similar mass but stronger feedback produce different results.
For example, the m11 simulation of the FIRE project \citep{Hopkins14} uses the same initial conditions.
However, its simulated galaxy resembles a dwarf spheroidal galaxy, inconsistent with the typical disc morphologies of isolated galaxies of that mass.
It is probable that the strong feedback in FIRE plus its extreme star-formation efficiency prevents any settling of the gas into a rotationally supported disc.
Therefore, the evolution of disc kinematics can potentially discriminate between different models of feedback.

A more quantitative comparison between different models of feedback is beyond the scope of this paper. We refer to other initiatives like the AGORA Galaxy Simulation Comparison Project \citep{Kim}.
This project aims to compare results from state-of-the-art gravito-hydrodynamics codes widely used in the numerical community.  
When different codes use a common subgrid physics for radiative cooling, star formation and feedback, they overall agree well, 
 regardless of the intrinsic differences in the numerical schemes used \citep{Kim16}.
 This experiment reassures that different models of feedback could be compared in the near future.

Finally, these conclusions are based on a handful of simulations. Large, statistically significant samples of simulations with different models of star-formation and feedback are needed in order to clarify 
these overall trends.

\section*{Acknowledgements}

We acknowledge stimulating discussions with Mark Krumholz.
The simulations were performed  at NASA Advanced Supercomputing (NAS) at NASA Ames Research Center.
This work has been partly funded by  the ERC Advanced Grant, STARLIGHT: Formation of the First Stars (project number 339177).
 JRP acknowledges support from grant HST-GO-12060.12-A. AD was partly supported by the grants ISF 124/12, I-CORE 
PBC/ISF 1829/12, BSF 2014-273, PICS 2015-18, and NSF AST-1405962.




\bibliographystyle{mnras}
\bibliography{disc4} 


\bsp	
\label{lastpage}
\end{document}